\documentclass[conference, a4paper]{IEEEtran}

\usepackage{balance}
\usepackage{cite}
\usepackage{amsmath,amssymb,amsfonts}
\usepackage{graphicx}
\usepackage{textcomp}
\usepackage{xcolor}
\def\BibTeX{{\rm B\kern-.05em{\sc i\kern-.025em b}\kern-.08em
    T\kern-.1667em\lower.7ex\hbox{E}\kern-.125emX}}

\usepackage{mathrsfs}
\usepackage{stfloats}
\usepackage{subfigure}
\usepackage{dsfont}
\usepackage{setspace}
\usepackage{array}
\usepackage{bbm}
\usepackage{hyperref}
\usepackage{bm}

\usepackage{accents}

\usepackage{xcolor}
\usepackage{epstopdf}

\usepackage{algorithm,algorithmic}

\allowdisplaybreaks[4]    

\begin{document}

\title{\huge Modeling the Mutual Coupling \\of Reconfigurable Metasurfaces

\author{\IEEEauthorblockN{
Marco Di Renzo\IEEEauthorrefmark{1},
Vincenzo Galdi\IEEEauthorrefmark{2}, and
Giuseppe Castaldi\IEEEauthorrefmark{2}
}                                     
\IEEEauthorblockA{\IEEEauthorrefmark{1}%
Universit\'e Paris-Saclay, CNRS, CentraleSup\'elec, Laboratoire des Signaux et Syst\`emes, 91192 Gif-sur-Yvette, France}
\IEEEauthorblockA{\IEEEauthorrefmark{2}%
University of Sannio,  Department of Engineering, 82100, Benevento, Italy}
\IEEEauthorblockA{\emph{marco.di-renzo@universite-paris-saclay.fr}}
}

}

\maketitle

\begin{abstract}
Recently, a circuits-based approach for modeling the mutual coupling of reconfigurable surfaces, which comprise sub-wavelength spaced passive scattering elements coupled with electronic circuits for enabling the reconfiguration of the surface, has been introduced. The approach is based on a finite-length discrete dipole representation of a reconfigurable surface, and on the assumption that the current distribution on each thin wire dipole is a sinusoidal function. Under these assumptions, the voltages at the ports of a multi-antenna receiver can be formulated in terms of the voltage generators at a multi-antenna transmitter through a transfer function matrix that explicitly depends on the mutual coupling and the tuning circuits through the mutual impedances between every pair of thin wire dipoles. In currently available works, the mutual impedances are formulated in an integral form. In this paper, we show that they can be formulated in a closed-form expression in terms of exponential integral functions.
\end{abstract}


\section{Introduction}
Dynamic (i.e., reconfigurable) metasurfaces have recently attracted major interest from the wireless research community \cite{MDR__JSAC_2020}, \cite{MDR__EURASIP_JIAN}, \cite{MDR__TCOM_THz}. These structures can be utilized for different applications in wireless networks, which include the implementation of smart reflectors, holographic transceivers, and single-RF multiple-input multiple-output modulators \cite{MDR__Elsevier_2022}. In spite of the many potential applications in wireless communications, the integration of dynamic surfaces into wireless systems needs to tackle many open challenges, which include the control overhead to enable the configuration of the surfaces every channel coherence interval \cite{MDR__JSTSP_2022}. From a communication-theoretic point of view, major open problems include the development of communication models for dynamic surfaces that are accurate enough but sufficiently tractable for performance analysis and optimization \cite{MDR__PIEEE_2022}, the development of efficient algorithms for optimization \cite{MDR__StefanTWC}, and the development of analytical frameworks to unveil fundamental performance trends and scaling laws \cite{MDR__XuewenWCL}, \cite{MDR__FadilSPAWC}, \cite{MDR__FadilTCOM}.

In general terms, a metasurface is defined as a reflectarray with sub-wavelength inter-distances between the array elements \cite{Sergei}, \cite{Vincenzo}. Sub-wavelength dynamic surfaces play an important role for controlling the wavefront of electromagnetic waves with a high power efficiency. In this context, a fundamental question is: \textit{how finely an electric field needs to be sampled in order to recover it exactly?} \cite{Book}. The answer lies in the sampling theorem applied to the wavenumber domain representation of the electromagnetic field under consideration. For the exact reconstruction of an electromagnetic field, it is known that the sampling period depends on the observation plane with respect to a reference plane, e.g., the plane that contains a given natural or digitally controllable scatterer. As sketched in Fig. 1, at distances from the reference plane where the evanescent waves (i.e., waves that attenuate exponentially along the $z$-axis) that constitute the electromagnetic field are negligible and can be ignored, a sampling period equal to half-wavelength is enough to perfectly recover the electromagnetic field. At distances from the reference plane where the evanescent waves that constitute the electromagnetic field are not negligible and cannot be ignored, however, a sampling period equal to half-wavelength is not enough to perfectly recover the electromagnetic field because the evanescent waves create rapid oscillations that cannot be reconstructed with a half-wavelength sampling period \cite[Section 1.4.2]{Book}. An illustration of the relation between the bandwidth of an electromagnetic field in the wavenumber domain and the observation plane (the distance with respect to the $xy$-plane in Fig. 1) is illustrated in \cite[Fig. 2]{Luca}. Even though the evanescent waves decay exponentially along the $z$-axis and are typically ignored in wireless communications, their control is, however, instrumental for realizing high-performance metasurfaces, e.g., perfect anomalous reflectors with high power efficiency and a large field of view (i.e., a large angle of reflection with respect to the angle of incidence) \cite{Sergei}.
\begin{figure}[!t]
\centering 
\centering\includegraphics[width=0.9\columnwidth]{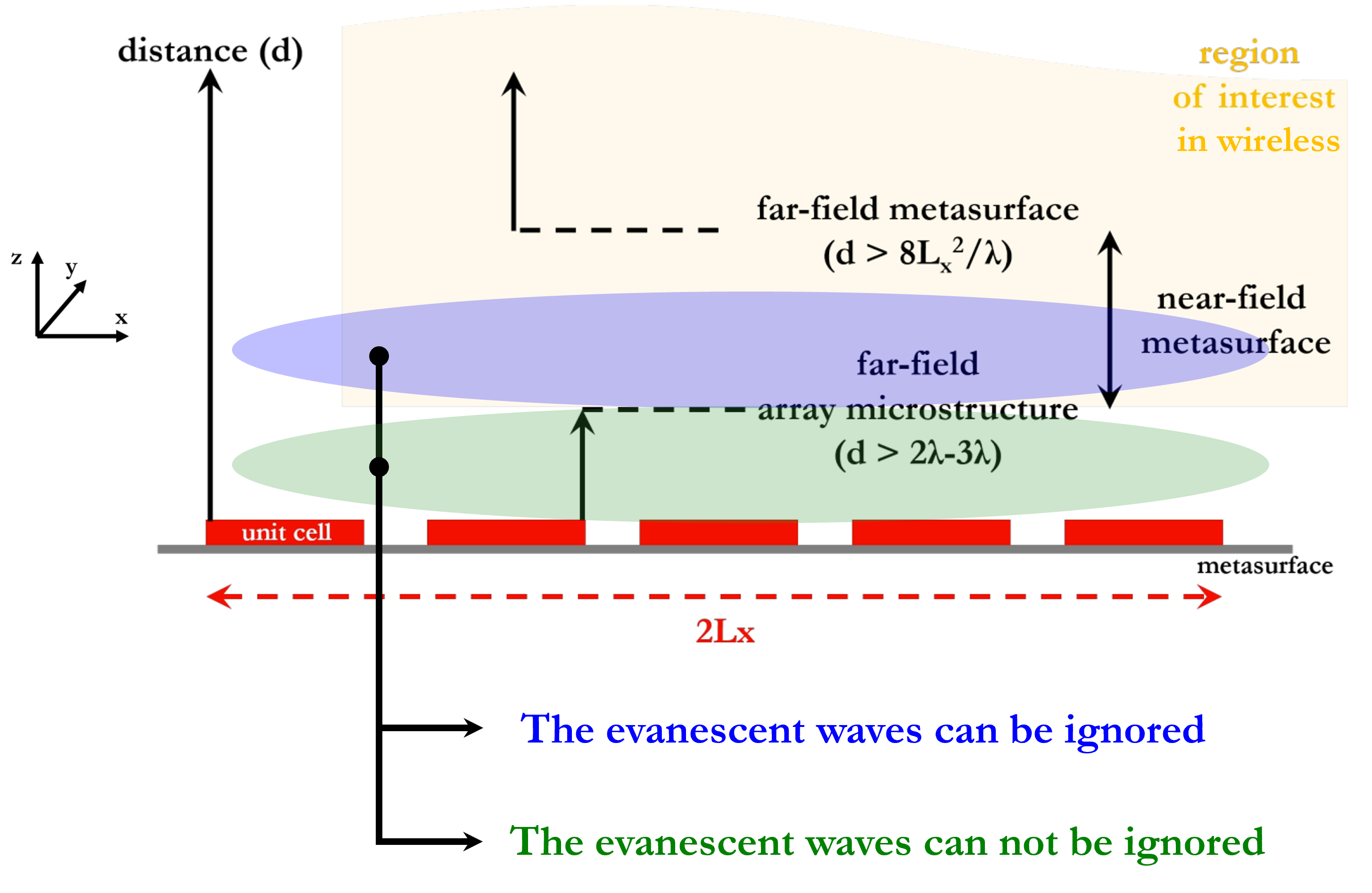}
\caption{Schematic illustration of the region of influence of evanescent waves.}
\end{figure}

Reconfigurable surfaces with sub-wavelength inter-distances between the scattering elements are, however, not easy to model and analyze. Due to the closely spaced scattering elements, the mutual coupling among them cannot be ignored and need to be taken in account at the design stage. It is known, in fact, that an appropriate optimization of the mutual coupling of antenna arrays may result in super-directive designs in which the gain scales with the square of the number of antenna elements, provided that the Ohmic losses are sufficiently low \cite{Marzetta}. In the context of modeling the mutual coupling of sub-wavelength reconfigurable surfaces, the authors of \cite{MDR_MC__Gabriele} have recently introduced a circuits-based approach for application to relay-type reconfigurable surfaces, which are known as reconfigurable intelligent surfaces (RIS), that explicitly accounts for the sub-wavelength spacing among the passive scattering elements and the circuital model for enabling their configuration. The approach is based on a finite-length discrete dipole representation \cite{DDA} for a reconfigurable surface and on the assumption that the current distribution on each thin wire dipole is a sinusoidal function. The authors of \cite{MDR_MC__Xuewen} and \cite{MDR_MC__Andrea} have subsequently shown that the performance of RIS-aided systems can be substantially improved if the mutual coupling is taken into account by design. In particular, mutual coupling aware designs enable either a better utilization of the aperture of the surface, since a larger number of densely spaced scattering elements can be deployed on the same surface area, or a reduction of the form factor of the surface, since the same number of densely spaced scattering elements can be deployed on a smaller surface area. This tradeoff is sketched in Fig. 2.

In \cite{MDR_MC__Gabriele}, the end-to-end transfer function matrix between a pair of transmitter and receiver is formulated in terms of mutual impedances that depend on the scattering elements and the tuning circuits for enabling the configuration of the surface. In \cite{MDR_MC__Gabriele}, however, the mutual impedances are formulated in an integral form. In this paper, we show that they can be formulated in a closed-form expression by utilizing a simple approach borrowed from \cite{Orfanidis}. Thanks to the obtained analytical expressions, the analysis and optimization of reconfigurable surfaces in the presence of mutual coupling is greatly simplified.

\begin{figure}[!t]
\centering 
\centering\includegraphics[width=0.80\columnwidth]{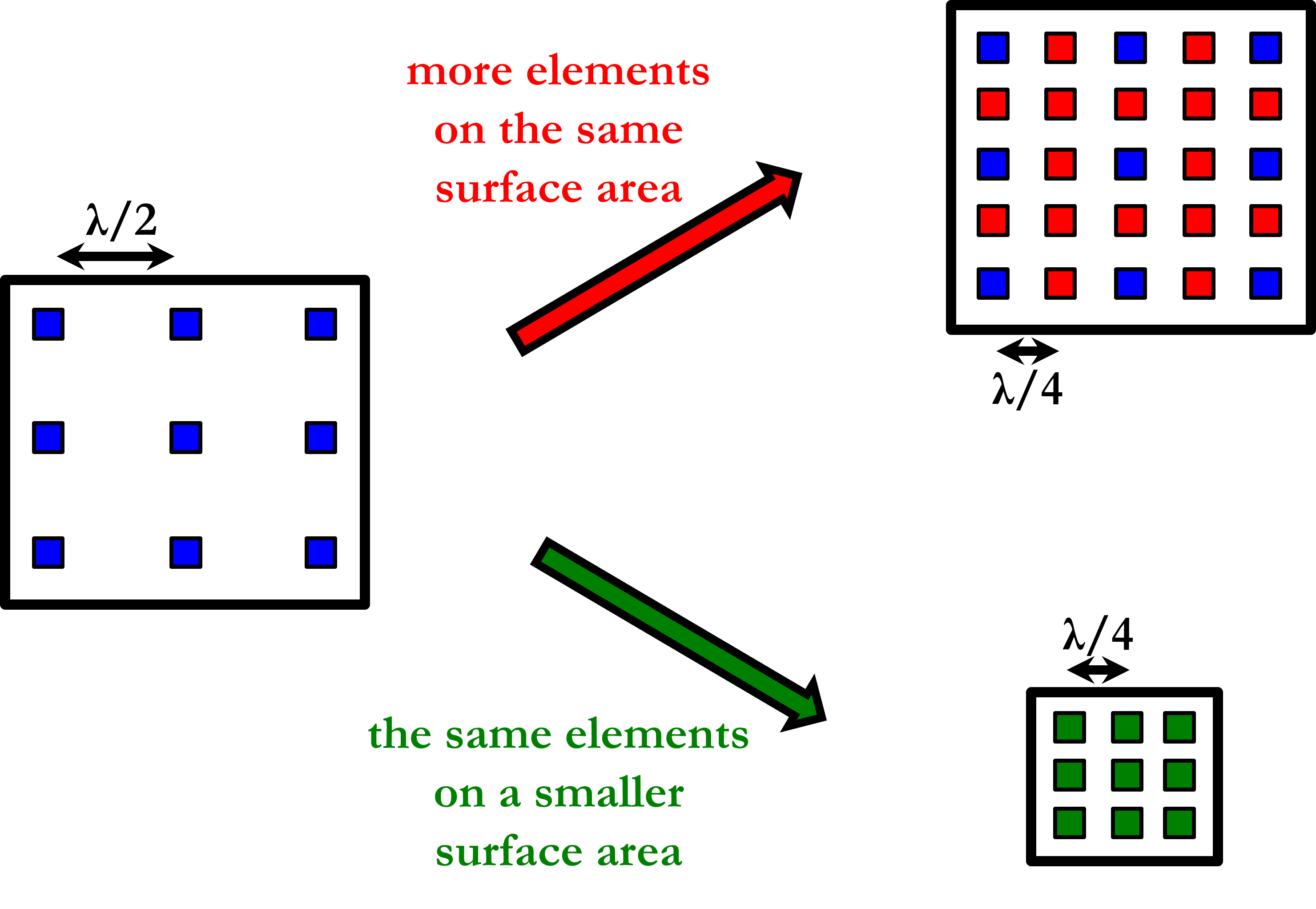}
\caption{Schematic illustration of the impact of sub-wavelength designs on utilizing a given surface area.}
\end{figure}
\begin{figure}[!t]
\centering 
\centering\includegraphics[width=0.98\columnwidth]{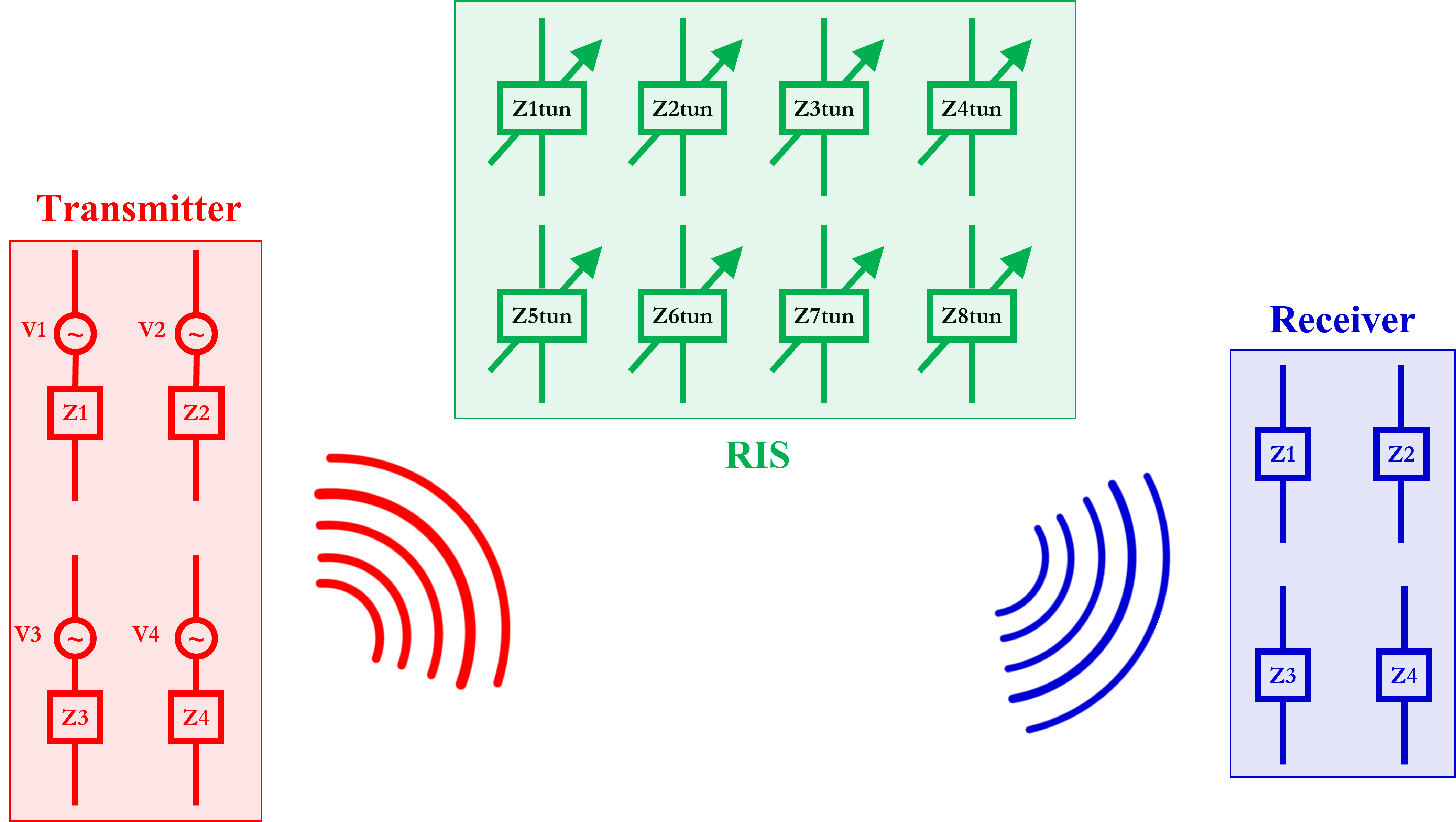}
\caption{Schematic illustration of the impedance-based model for relay-type surfaces introduced in \cite{MDR_MC__Gabriele}, \cite{MDR__PIEEE_2022}.} \vspace{-0.5cm} 
\end{figure}
\section{Closed-Form Expression of the Mutual Impedances}
For simplicity, we depart from \cite[Corollary 1]{MDR_MC__Gabriele} that is applicable to single-antenna transmitters and single-antenna receivers under the assumption that they are in the far-field of each scattering element that constitutes an electrically large reconfigurable surface made of $N$ reconfigurable elements. Therefore, the model is applicable to the near-field and the far-field regions of the whole surface. For self-consistency, a schematic illustration of the impedance-based model introduced in \cite{MDR_MC__Gabriele} is reported in Fig. 3 \cite{MDR__PIEEE_2022}.

The end-to-end channel can be written as follows\footnote{$(\cdot)^T$ denotes the transpose operator.}:
\begin{equation}
{h_{e2e}} = {z_{{\rm{RT}}}} - {\bf{z}}_{{\rm{RS}}}^T{\left( {{{\bf{Z}}_{{\rm{SS}}}} + {{\bf{Z}}_{{\rm{RIS}}}}} \right)^{ - 1}}{{\bf{z}}_{{\rm{ST}}}}
\end{equation}
where ${z_{{\rm{RT}}}}$ is the mutual impedance between the transmitter and the receiver, ${\bf{z}}_{{\rm{RS}}}$ is the $N \times 1$ vector of mutual impedances between the $N$ reconfigurable elements of the surface and the receiver, ${\bf{z}}_{{\rm{ST}}}$ is the $N \times 1$ vector of mutual impedances between the transmitter and the $N$ reconfigurable elements of the surface, ${{{\bf{Z}}_{{\rm{RIS}}}}}$ is the $N \times N$ diagonal matrix containing the $N$ equivalent impedances of the tuning circuits enabling the reconfigurability of the surface, and ${{{\bf{Z}}_{{\rm{SS}}}}}$ is the $N \times N$ (full) matrix containing the mutual impedances between every pair of scattering elements of the surface. Under the modeling assumptions in \cite{MDR_MC__Gabriele}, ${{{\bf{Z}}_{{\rm{SS}}}}}$ is independent of ${{{\bf{Z}}_{{\rm{RIS}}}}}$. 

We are interested in computing the generic element of ${{{\bf{Z}}_{{\rm{SS}}}}}$. For ease of writing, let us denote the $(qp)$th element of ${{{\bf{Z}}_{{\rm{SS}}}}}$ by ${z_{qp}}$. Specifically, ${z_{qp}}$ denotes the contribution of the electric field generated by the current flowing through the $p$th reconfigurable element of the surface when it is observed on the $q$th reconfigurable element of the surface. From \cite{MDR_MC__Gabriele}, we have the following:
\begin{equation}
{z_{qp}} = \frac{{j\eta }}{{4\pi k}}\int\nolimits_{ - {h_q}}^{ + {h_q}} {{E_{qp}}\left( z \right){f_q}\left( z \right)dz}
\end{equation}
where
\begin{equation}
{E_{qp}}\left( z \right) = \left( {\partial _z^2 + {k^2}} \right)\int\nolimits_{ - {h_p}}^{ + {h_p}} {\left( {\frac{{\exp \left( { - jk{R_{qp}}\left( {\xi ,z} \right)} \right)}}{{{R_{qp}}\left( {\xi ,z} \right)}}} \right){f_p}\left( \xi  \right)d\xi }
\end{equation}
with
\begin{equation}
{f_p}\left( \xi \right) = \frac{{\sin \left( {k\left( {{h_p} - \left| \xi  \right|} \right)} \right)}}{{\sin \left( {k{h_p}} \right)}},\quad {f_q}\left( z \right) = \frac{{\sin \left( {k\left( {{h_q} - \left| z \right|} \right)} \right)}}{{\sin \left( {k{h_q}} \right)}}
\end{equation}
and
\begin{equation}
\begin{array}{l}
{R_{qp}}\left( {\xi ,z} \right) = \sqrt {d_{qp}^2 + {{\left( {{z_q} - {z_p} + z - \xi } \right)}^2}} \quad {\rm{if}}\quad p \ne q\\
{R_{qp}}\left( {\xi ,z} \right) = \sqrt {a_q^2 + {{\left( {z - \xi } \right)}^2}} \quad {\rm{if}}\quad p = q
\end{array}
\end{equation}
where $d_{qp}^2 = {\left( {{x_q} - {x_p}} \right)^2} + {\left( {{y_q} - {y_p}} \right)^2}$. In (2)-(5), the following notation is utilized: $h_p$ and $h_q$ denote the half-length of the $p$th and $q$th reconfigurable element, respectively; $a_q$ denotes the radius of the $q$th reconfigurable element; $(x_p, y_p, z_p)$ and $(x_q, y_q, z_q)$ are the center locations of the $p$th and $q$th reconfigurable element, respectively, which are all aligned along the $z$-axis; $k=2\pi/\lambda$ where $\lambda$ is the wavelength; $\eta$ is the intrinsic impedance of free space; $j$ is the imaginary unit; and ${\partial _z^2}$ is the second-order partial derivative with respect to $z$.

The computation of ${z_{qp}}$ requires the calculation of the electric field in (3) and then the integral in (2).

\subsection{Computation of (3)}
The electric field in (3) can be formulated in a closed-form expression by applying the approach proposed in \cite[Section 25.1]{Orfanidis}. The final result is the following:
\begin{equation}
\begin{split}
{E_{qp}}\left( z \right) &= \frac{k}{\sin\left( {k{h_p}} \right)}\frac{{\exp \left( { - jk{R_{qp}}\left( {\xi  =  + {h_p},z} \right)} \right)}}{{{R_{qp}}\left( {\xi  =  + {h_p},z} \right)}}\\
& + \frac{k}{\sin\left( {k{h_p}} \right)}\frac{{\exp \left( { - jk{R_{qp}}\left( {\xi  =  - {h_p},z} \right)} \right)}}{{{R_{qp}}\left( {\xi  =  - {h_p},z} \right)}}\\
& - \frac{2k\cos \left( {k{h_p}} \right)}{\sin\left( {k{h_p}} \right)}\frac{{\exp \left( { - jk{R_{qp}}\left( {\xi  = 0,z} \right)} \right)}}{{{R_{qp}}\left( {\xi  = 0,z} \right)}}
\end{split}
\end{equation}

\subsection{Computation of (2)}
Given the closed-form expression of the electric field in (6), ${z_{qp}}$ can be formulated in terms of exponential integral functions. Specifically, we first rewrite the function ${f_q}\left( z \right)$ as follows:
\begin{equation}
\begin{split}
{f_q}\left( z \right) &= \frac{1}{{2j\sin \left( {k{h_q}} \right)}} \\ &\times \sum\limits_{{s_0} = \left\{ { - 1, + 1} \right\}} {{s_0}\exp \left( {j{s_0}k{h_q}} \right)\exp \left( { - j{s_0}k\left| z \right|} \right)} 
\end{split}
\end{equation}

Then, ${z_{qp}}$ can be rewritten as follows:
\begin{equation}
\begin{split}
{z_{qp}} &= \frac{{\eta {c_q}}}{{8\pi }}\sum\limits_{{s_0} = \left\{ { - 1, + 1} \right\}} {{s_0}\exp \left( {j{s_0}k{h_q}} \right){I_{qp}}\left( {{\xi _p} = +{h_p};{s_0}} \right)} \\
& + \frac{{\eta {c_q}}}{{8\pi }}\sum\limits_{{s_0} = \left\{ { - 1, + 1} \right\}} {{s_0}\exp \left( {j{s_0}k{h_q}} \right){I_{qp}}\left( {{\xi _p} =  - {h_p};{s_0}} \right)} \\
& - \frac{{\eta {c_{qp}}}}{{8\pi }}\sum\limits_{{s_0} = \left\{ { - 1, + 1} \right\}} {{s_0}\exp \left( {j{s_0}k{h_q}} \right){I_{qp}}\left( {{\xi _p} = 0;{s_0}} \right)} 
\end{split}
\end{equation}
where ${c_q} = {1 \mathord{\left/ {\vphantom {1 {\sin \left( {k{h_q}} \right)}}} \right. \kern-\nulldelimiterspace} {\sin \left( {k{h_q}} \right)}}$, ${c_{qp}} = 2\cos \left( {k{h_p}} \right){c_q}$, and
\begin{equation}
\begin{split}
&{I_{qp}}\left( {{\xi _p};{s_0}} \right) \\ &= \int\nolimits_{ - {h_q}}^0 {\frac{{\exp \left( { - jk\sqrt {\rho _{qp}^2 + {{\left( {z + z_{qp} - {\xi _p}} \right)}^2}} } \right)}}{{\sqrt {\rho _{qp}^2 + {{\left( {z + z_{qp} - {\xi _p}} \right)}^2}} }}\exp \left( {j{s_0}kz} \right)dz} \\
& + \int\nolimits_0^{ + {h_q}} {\frac{{\exp \left( { - jk\sqrt {\rho _{qp}^2 + {{\left( {z + z_{qp} - {\xi _p}} \right)}^2}} } \right)}}{{\sqrt {\rho _{qp}^2 + {{\left( {z + z_{qp} - {\xi _p}} \right)}^2}} }}\exp \left( { - j{s_0}kz} \right)dz} 
 \end{split}
\end{equation}
with $\rho _{qp}^2 = d_{qp}^2$ if $p \ne q$ and $\rho _{qp}^2 = a_q^2$ if $p=q$, respectively, and $z_{qp} = z_q - z_p$.

The integral in (9) can be formulated in terms of exponential integral functions by applying the change of variables in \cite[Appendix G]{Orfanidis}. Specifically, let us consider the following notable integral:
\begin{equation}
\begin{split}
\mathcal{J}\left(s_0, d_0, z_0; L, U\right) \\& \hspace{-2.5cm} = \int\nolimits_L^U {\exp \left( { - jk{s_0}\zeta } \right)\frac{{\exp \left( { - jk\sqrt {d_0^2 + {{\left( {\zeta  - {z_0}} \right)}^2}} } \right)}}{{\sqrt {d_0^2 + {{\left( {\zeta  - {z_0}} \right)}^2}} }}d\zeta } \\
& \hspace{-2.5cm}= {s_0}\exp \left( { - jk{s_0}{z_0}} \right){E_1}\left( {jk{L_0}} \right)\\
& \hspace{-2.5cm} - {s_0}\exp \left( { - jk{s_0}{z_0}} \right){E_1}\left( {jk{U_0}} \right)
\end{split}
\end{equation}
where 
\begin{equation}
\begin{array}{l}
{L_0} = \sqrt {d_0^2 + {{\left( {L - {z_0}} \right)}^2}}  + {s_0}\left( {L - {z_0}} \right)\\
{U_0} = \sqrt {d_0^2 + {{\left( {U - {z_0}} \right)}^2}}  + {s_0}\left( {U - {z_0}} \right)
\end{array}
\end{equation}
and
\begin{equation}
{E_1}\left( c \right) = \int\nolimits_c^\infty  {\frac{{\exp \left( { - u} \right)}}{u}du}
\end{equation}
is the exponential integral function with $c$ being a complex number whose phase is such that $\left| {\arg \left( c \right)} \right| < \pi$.

Then, ${\mathcal{I}_{qp}}\left( {{\xi _p};{s_0}} \right)$ can be formulated as follows:
\begin{equation} 
\begin{split}
{{\mathcal{I}}_{qp}}\left( {{\xi _p};{s_0}} \right) &= {\mathcal{J}}\left( { - {s_0},{\rho _{qp}},{\xi _p} - {z_{qp}}; - {h_q},0} \right)\\
& + {\mathcal{J}}\left( { + {s_0},{\rho _{qp}},{\xi _p} - {z_{qp}};0, + {h_q}} \right)
\end{split}
\end{equation}

The integral in (13) can be applied to any setup provided that $\rho_{qp} \ne 0$, because $E_1(c)$ is not defined for $c=0$. If $\rho_{qp} = 0$, e.g., the scattering elements are in a collinear formation\cite[Fig. 8.20(b)]{Proakis}, a specific closed-from expression is available \cite[Eq. (8-72)]{Proakis}. Alternatively, (9) can be directly computed numerically.

\section{Conclusion} 
In this paper, we have discussed the engineering relevance of considering reconfigurable surfaces whose scattering elements are spaced more densely than half-wavelength, in order to model the presence of evanescent waves in the close vicinity of the surface, and to be able to control them for realizing extreme manipulations of the electromagnetic waves with high power efficiency. In these implementations, the mutual coupling among the scattering elements cannot be ignored. Departing from the approach recently introduced in \cite{MDR_MC__Gabriele} for modeling the mutual coupling among closely-spaced scattering elements, we have shown that the end-to-end channel transfer function matrix can be formulated in a closed-form expression in terms of exponential integral functions, which are built-in functions in state-of-the-art programming and numeric computing platforms. This greatly simplifies the analytical calculation of the end-to-end channel transfer function matrix in \cite{MDR_MC__Gabriele}, and the corresponding optimization of the tuning elements of the surface for maximizing the system performance \cite{MDR_MC__Xuewen}, \cite{MDR_MC__Andrea}.

\section*{Acknowledgment}
This work was supported by the European Commission through the H2020 ARIADNE (grant 871464) and RISE-6G (grant 101017011) projects, and the Fulbright Foundation.

\end{document}